\documentclass[12pt]{article}

\usepackage{amssymb}
\usepackage{amsmath}

\usepackage{amsbsy}

\usepackage{rotating}

\usepackage[usenames]{color} 
\usepackage{colortbl} 

\usepackage{graphics}
\usepackage{epsfig}

\begin{document}%

 \title{Dilaton generation during the propagation of magnetic dipole radiation of a rotating neutron star through the Coulomb field of charged particle}

\author{M.O.\, Astashenkov$^{a, b}$\footnote{E-mail: mixa.astash@yandex.ru}, P.A.Vshivtseva$^{a, b}$, E.T.\, Einiev$^{c}$ \\ \\
{\small $^{a}$\,Department of Physics, Lomonosov Moscow State University, 119991, Moscow,  Russia} \\
{\small $^{b}$\, Skobeltsyn Institute of Nuclear Physics, Lomonosov Moscow State University, 119991,}\\ {\small Moscow, Russia}\\
{\small $^{c}$\, National Research University Moscow Power Engineering Institute, 111250, Moscow, Russia}}

\maketitle 

\thispagestyle{empty}

\bigskip
\begin{abstract}
According to the Dilaton-Einstein-Maxwell theory, dilatons can be generated by electromagnetic fields with a non-zero the Maxwell tensor invariant.
This research focuses on the dilaton generation while magnetic dipole radiation from a rotating neutron star propagates through a Coulomb field created by a point-charged particle.

The exact solution of the dilaton field equation has been obtained, and the angular distribution of dilaton generation has been established.

\end{abstract}

\section*{Introduction}

A lot of studies have focused on examining the characteristics of heavy and light axion-like particles that exist beyond the Standard Model and are generated by different electromagnetic fields and waves:  arions
\cite{arions-from-pulsar, arions-gltc-magn-fld, aroin-3}, axions \cite{axion-1, axion-2, axion-3} and dilatons \cite {dilaton-1, dil-from-pulsar, dil-Q-plane_wave}. According to modern concepts, arions and dilations are massless particles, whereas axions have mass.

The reason for studying dilatons is driven by multidimensional theories of gravity such as superstring \cite{string-th} theory and the 5-dimensional Kaluza-Klein theory \cite{Kaluza-klein}.

Electromagnetic fields and waves as sources of
axion-like particles were theoretically considered as in laboratory\cite {ALP-exp1,	ALP-exp2}, and in astrophysical conditions. Specifically, the research \cite{dil-Q-plane_wave} analyzes the dilaton generation while a plane electromagnetic wave propagates through the Coulomb field of a point-charged particle. The study \cite{dil-Q-plane_wave} has shown that dilaton waves are produced at the identical frequency as the electromagnetic wave. In astrophysical conditions, it is interesting to consider coherent electromagnetic radiation propagating in the interstellar medium through charged particles as the source of dilaton generation.

As the source of coherent electromagnetic waves, one can consider rotating neutron stars in which the magnetic dipole momentum's direction differs from the axis of rotation. As it's known, neutron stars are compact objects with magnetic fields that range from $10^{8}$ to $10^{15}$ Gauss. 

For this reason, this study is focused on dilaton generation during propagation of magnetic dipole radiation of a rotating neutron star in the Coulomb field of charged particle from the interstellar medium.

\section{Dilaton field equations}
According to the study \cite {dil-from-pulsar}, the Lagrange function density for the dilaton field,
interacting with the electromagnetic field, can be expressed in the form:

\begin{equation} \label{L}
	{\cal L} =a_0(\partial\Psi)^2+a_1e^{-2{\cal K}\Psi}F_{nm}F^{nm},
\end{equation}
where $a_0,\ a_1$ and $\cal K$ are gauge constant, $(a_0>0)$, 
$\Psi$ is a scalar field of dilaton, and $F_{nm}F^{nm}$ is the invariant of the Maxwell tensor $F_{nm}$.

The value of $ {\cal K} = 1 $ is predicted in string theory, while for 5-dimensional Kaluza-Klein theory: $ {\cal K} =\sqrt {3} $. In this paper, the parameter $ {\cal K} $ is considered arbitrary.

The field equations resulting from the Lagrange function density \eqref{L} in pseudorimanian space take on the following structure:
\begin{equation}
	(\Delta - \frac{1}{c^2}\frac{\partial^2}{\partial t^2}) \Psi={a_1{\cal K}\over a_0}e^{-2{\cal K}\Psi} F_{nm}F^{nm}
	={2a_1{\cal K}\over a_0} 
	e^{-2{\cal K}\Psi}\big[{\bf B}^2-{\bf E}^2\big],
\end{equation}
where $\bf E$ and $\bf B$ are the electric field strength and magnetic induction correspondingly.

Since dilatons have not been discovered, one can assume that the dilaton field is sufficiently weak that the following inequality is satisfied: $|{\cal K }\Psi| \ll 1$. In this scenario, the dilaton field equation will take the form:
\begin{equation} \label{dil_eq_b2}
	(\Delta - \frac{1}{c^2}\frac{\partial^2}{\partial t^2}) \Psi=\frac{2a_1{\cal K}}{a_0} \big[{\bf B}^2-{\bf E}^2\big].
\end{equation}
Under the approximation being considered, it can be inferred from equation \eqref{dil_eq_b2} that electromagnetic fields with a non-zero invariant ${\bf B}^2 - {\bf E}^2$ can potentially be sources of dilaton generation.

\section{Problem statement}
Suppose that a point particle with charge Q is placed at the beginning of the Cartesian coordinate system. The static field $E_0 $, created by the charge $Q$, in the Maxwell approximation has the form:
\begin{equation} \label{E_0}
	{\bf E}_0=\frac{Q{\bf r}}{r^3}.
\end{equation}
Suppose there exists a neutron star (such as a pulsar or magnetar) rotating at a frequency of $\omega$ around the $z$-axis with a radius of $R_S$, located at coordinates ${\bf r}_0 =\{ x_0, y_0, z_0\}$ where $R_S$ is less than $r_0$.
The magnetic dipole momentum ${\bf m} =\{ m_x, m_y, m_z\}$ is oriented at an angle $\alpha $ with respect to the $z$-axis. In this situation, the magnetic dipole moment ${\bf m}$ can be expressed in the following manner:
\begin{equation} \label{m}
	{\bf m}(t)= m(\cos(\omega t)\sin\alpha, \ 
	\sin(\omega t)\sin\alpha,\ \cos\alpha),
\end{equation}
where $m = |{\bf m}|$ is the modulus of the magnetic dipole momentum vector ${\bf m}$.

The magnetodipole radiation of a neutron star will take the form:
\begin{equation} \label{E_w}
	{\bf E}_w({\bf r},t)={({\bf R}\times\dot{\bf m}(\tau))\over c R^3}+
	{({\bf R}\times\ddot{\bf m}(\tau))\over c^2 R^2},
\end{equation}
\begin{align} \label{B_w}
	{\bf B}_w({\bf r},t)
	={3({\bf m}(\tau)\cdot {\bf R}){\bf R}
		-R^2{\bf m}(\tau)\over  R^5}
	-{{ \dot{\bf m}}(\tau)\over c R^2}+\nonumber \\
	+{3({ \dot{\bf m}}(\tau)\cdot {\bf R}){\bf R}\over c R^4}
	+{(\ddot{\bf m}(\tau)\cdot {\bf R}){\bf R}-R^2\ddot{\bf m}(\tau)
		\over c^2 R^3},
\end{align}
where ${\bf R}={\bf r-r}_0$. In the expressions\eqref {E_w} ,\eqref{B_w}, the magnetic dipole momentum ${\bf m} $ depends on the retarded time $\tau = t - R/c $, and the point above the vector denotes derivative by $\tau$.

In general, a neutron star has a gravitational field. As metric of a pseudo-rimmed space tensor, we will use the Schwarschild solution \cite{Schwarzschild} in isotropic coordinates \cite{landau}:
\begin{equation*}
	g_{00}={(4R-r_g)^2\over (4R+r_g)^2},\ \ \ \ \ \
	g_{xx}=g_{yy}=g_{zz}=-(1+{r_g\over 4R})^4,
\end{equation*}
where $r_g$ is Schwarzschild radius. Since the typical radius of a neutron star is $R_S\sim 10$ km \cite{NS_rad}, and the mass of $ M_S $ is estimated from $0.1 $ to $1 $ solar masses, the ratio $r_ g/R_S$ can be estimated to be $\sim 0.01$. 
Because the point charge $Q$ is located outside the neutron star in the scenario being studied, the inequality $r_0 > R_S > r_g$ is satisfied, leading to $r_g/r_0 < 0.01$.
Thus, the impact of gravitational field on magnetic dipole radiation\eqref {E_w} ,\eqref {B_w} is minimal, and in the first approximation, the Minkowski metric tensor can be used as the metric tensor: $g_{00} = 1,\; g_{11}=g_{22} = g_{33} = -1$.

In the case under consideration, the time-dependent part of the first invariant of the electromagnetic field tensor has the form:
\begin{equation} \label{F_inv}
	F_{nm}F^{nm}=2\big[{\bf B}^2-{\bf E}^2\big]=
	2\big[{\bf B}_w^2-{\bf E}_w^2\big]-4({\bf E}_0\cdot{\bf E}_w).
\end{equation}
The first two terms on the right side of this equation come from the electromagnetic field of the rotating magnetic dipole of the neutron star\eqref {E_w} ,\eqref {B_w} and have been studied in detail in the research \cite{dil-from-pulsar}.
According the study in \cite{dil-from-pulsar} the generation of dilaton has been shown to take place at the frequencies $\omega$ and $2\omega$. In this paper, we will focus on the study of the dilaton generation in the Coulomb field of charged particle, which is provided by the term $({\bf E}_0\cdot {\bf E}_w)$ in the expression \eqref{F_inv}.

\section{Exact solution search}
By replacing equation \eqref{F_inv} in equation \eqref{dil_eq_b2}, one can derive the subsequent equation:
\begin{equation} \label{Q_psi_eq_E0Ew}
	(\Delta - \frac{1}{c^2}\frac{\partial^2}{\partial t^2}) \Psi=-\frac{4a_1{\cal K}}{a_0}({\bf E}_0\cdot{\bf E}_w).
\end{equation}
The equation \eqref{Q_psi_eq_E0Ew} describes dilaton generation during the propagation of magnetic dipole radiation \eqref{E_w} and \eqref{B_w} through the Coulomb field of an electrically charged particle \eqref{E_0}.

Substituting expressions for ${\bf E}_0$ \eqref{E_0} and ${\bf E}_w$ \eqref{E_w} into equation \eqref{Q_psi_eq_E0Ew}, one can derive the following equation:
\begin{align} \label{Q_psi_main_eq}
	(\Delta - \frac{1}{c^2}\frac{\partial^2}{\partial t^2}) \Psi=-\frac{4a_1{\cal K}Q}{a_0r^3}\Big[\frac{\big({\bf r} \cdot ({\bf R} \times \dot{{\bf m}})\big)}{cR^3} \nonumber \\
	+\frac{\big({\bf r}\cdot ({\bf R} \times \ddot{\bf m})\big)}{c^2R^2}\Big]\;.
\end{align}

Note that the expression \eqref{E_w} can be represented in following manner:
\begin{equation*}
	{\bf E}_w = \mathrm{rot}_{r_0} \frac{\dot{{\bf m}}(\tau)}{cR}\;,
\end{equation*}
where $\mathrm{rot}_{r_0} $ is the $\mathrm{rot}$ operator that performs differentiation using the coordinates of the vector ${\bf r}_0$, instead of ${\bf r}$. Following this, the equation \eqref{Q_psi_main_eq} can be expressed in the following manner:
\begin{align} \label{Q_psi_eq1}
	(\Delta - \frac{1}{c^2}\frac{\partial^2}{\partial t^2}) \Psi=\frac{4a_1{\cal K}Q}{a_0r^3} \;\mathrm{div}_{r_0} ({\bf r} \times \frac{\dot{{\bf m}}(\tau)}{cR})\;.
\end{align}
One possible way to find a solution to the equation \eqref{Q_psi_eq1} is as follows:
\begin{equation} \label{Q_psi_F}
	\Psi({\bf r}, t) = \frac{4a_1{\cal K}Q}{a_0}\; \mathrm{div}_{r_0} {\bf F}({\bf r}, t)\;.
\end{equation}
By replacing the value of \eqref{Q_psi_F} in the equation \eqref{Q_psi_eq1}, one can derive the subsequent equation for the unidentified vector function ${\bf F}$:
\begin{equation} \label{Q_eq_for_F1}
	(\Delta - \frac{1}{c^2}\frac{\partial^2}{\partial t^2}) {\bf F}({\bf r}, t) = \frac{1}{r^3}({\bf r} \times \frac{\dot{{\bf m}}(\tau)}{cR})\;.
\end{equation}

Rewrite the time-dependent part of the magnetic dipole momentum ${\bf m}$ \eqref{m} in a complex form. Then the derivative of the magnetic dipole moment $\dot{\bf m}$ can be written as follows:
\begin{align} \label{vec_M}
	\dot{\bf m}(\tau) = -i\omega m\sin\alpha \{1, i, 0\} \frac{e^{-i(\omega t - kR)}}{R} \nonumber \\= -i\omega \; {\bf M} \frac{e^{-i(\omega t - kR)}}{R}\;,
\end{align}
where the notation ${\bf M} = m\sin\alpha\{1, i, 0\}$ is defined.
Taking this into consideration, the equation\eqref {Q_eq_for_F1} will be expressed as:
\begin{equation} \label{Q_eq_for_F2}
	(\Delta - \frac{1}{c^2}\frac{\partial^2}{\partial t^2}) {\bf F}({\bf r}, t) = \frac{-i\omega}{cr^3}({\bf r} \times {\bf M})\frac{e^{-i(\omega t - kR)}}{R}\;.
\end{equation}
The unknown vector function ${\bf F}({\bf r}, t) $ can be discovered in the form:
\begin{equation} \label{F_and_P}
	{\bf F}({\bf r}, t) =\frac{i\omega}{c}({\bf P}({\bf r}, t) \times {\bf M})\;.
\end{equation}
By taking into account equations \eqref {Q_eq_for_F2} and \eqref {F_and_P}, one can derive the following equation for the new unknown vector function ${\bf P}$:
\begin{equation} \label{eq_for_P}
	(\Delta - \frac{1}{c^2}\frac{\partial^2}{\partial t^2}) {\bf P}({\bf r}, t) = - \frac{e^{-i(\omega t - kR)}}{r^3R}\;{\bf r}\;.
\end{equation}
The initial task is to locate a particular solution ${\bf P}_p$ of the equation \eqref {eq_for_P}. To achieve this, write the function ${\bf P}$ in the following manner:
\begin{equation}
	{\bf P}({\bf r}, t) = {\bf q}({\bf r})\frac{e^{-i(\omega t - kR)}}{R}\;.
\end{equation}
Next, the equation for the unknown function ${\bf q}$ can be derived:
\begin{equation} \label{eq_for_q}
	\Delta {\bf q} + \frac{2}{R}(ik - \frac{1}{R})({\bf R}\cdot \nabla){\bf q} = - \frac{{\bf r}}{r^3}\;.
\end{equation}
Exact solution of the equation \eqref{eq_for_q}:
\begin{equation}
	{\bf q} = \frac{i}{2k}[\frac{{\bf r}_0}{r_0}+\frac{{\bf r}}{r}\big]\frac{R}{r_0r+({\bf r}_0 \cdot {\bf r})}\;.
\end{equation}

The corresponding exact partial solution ${\bf P}_p$ to the equation \eqref {eq_for_P} has been identified as:
\begin{equation} \label{P_partial_sol}
	{\bf P}_p({\bf r}, t) = \frac{i}{2k}e^{-i\omega t}\big[\frac{{\bf r}_0}{r_0}+\frac{{\bf r}}{r}\big]\;\frac{e^{-i(\omega t -kR)}}{r_0r +({\bf r}_0 \cdot {\bf r})}\;.
\end{equation}
Denote the angle between the vectors ${\bf r_0}$ and $ {\bf r}$ as $\gamma$. Then the dot product between them is equal: $({\bf r}_0\cdot {\bf r}) = r_0r\cos\gamma$.
The obtained particular solution \eqref{P_partial_sol} has a singularity at $\cos\gamma\to -1$, even though the right side of the equation \eqref{eq_for_P} does not have the same singularity.
For this reason, one should include the solution $ {\bf P}_0$ of the homogenous equation \eqref {eq_for_P} to ensure that the overall solution $ {\bf P} = {\bf P}_0 + {\bf P}_p $ remains nonsingular as $\cos\gamma\to -1$.
As a result, one can obtain:
\begin{equation} \label{P_sol}
	{\bf P}({\bf r}, t) = \frac{i}{2k}e^{-i\omega t}\big[\frac{{\bf r}_0}{r_0}+\frac{{\bf r}}{r}\big]\;\frac{e^{ikR} - e^{ik(r_0+r)}}{r_0r +({\bf r}_0 \cdot {\bf r})}
\end{equation}

Finally, by utilizing expressions\eqref{Q_psi_F}, \eqref{F_and_P}, \eqref{P_sol}, one can derive an exact solution to the dilaton field equation \eqref{Q_psi_main_eq}:
\begin{align} \label{Q-psi_sol1}
	\Psi = -\frac{2i {\cal K} a_1 Q k}{a_0(r_0r+({\bf r_0 \cdot r}))} e^{-iwt}\big[\frac{1}{R}(\frac{1}{r_0}+\frac{1}{r})e^{ikR} \nonumber \\- \frac{1}{r_0r}e^{ik(r_0+r)}\big] \big({\bf M} \cdot [{\bf r_0}\times {\bf r}]\big)\;.
\end{align}

Specify the vector ${\bf r}_0$ that indicates the location of the charged point particle, by using the azimuth angle $\theta_0$ and the polar angle $\phi_0$:
\begin{equation} \label{r_0}
	{\bf r}_0 = r_0\{\sin\theta_0\cos\phi_0, \; \sin\theta_0\sin\phi_0, \; \cos\theta_0\}\;.
\end{equation}
After that, in spherical coordinates, the dilaton field $\Psi$ \eqref{Q-psi_sol1} will appear as:
\begin{align} \label{Q-psi-sphr}
	\Psi = \frac{2{\cal K}a_1 kQ m \sin\alpha}{a_0 r_0 r (1+\cos\gamma)}e^{-iwt}\big[\frac{r_0+r}{R}e^{ikR} - e^{ik(r_0+r)}\big] \nonumber \\ \times\big(\cos\theta_0\sin\theta e^{i\phi}- \sin\theta_0 \cos\theta e^{i\phi_0}\big)\;.
\end{align}

In the given spherical coordinate system, the cosine of the angle between ${\bf r}_0$ and ${\bf r}$ is defined as follows:
\begin{equation*} \label{cos_gamma}
	\cos\gamma = \sin\theta_0\sin\theta\cos(\phi-\phi_0) + \cos\theta_0 \cos\theta\;.
\end{equation*} 

It is interesting to note that with $\theta \to \pi - \theta_0, \phi \to \phi_0+\pi$ as well as $\theta \to \theta_0, \phi \to \phi_0$ dilaton field $\Psi$ in the expression \eqref{Q-psi-sphr} refers to $0$.

\section{Angular distribution of dilaton generation}
Based on the Lagrange function density \eqref{L}, the stress-energy tensor $T^{ik}$ takes the following form for free dilaton field $\Psi$:
\begin{equation} \label{T_ik}
	T^{ik}=2a_0g^{in}g^{km} \big\{{\partial \Psi\over \partial x^n}
	{\partial \Psi\over \partial x^m}
	-{1\over 2}g_{mn}\frac{\partial \Psi}{ \partial x^s}
	\frac{\partial \Psi}{\partial x^q} g^{sq}\big\}.
\end{equation}

The energy density of the dilaton field $T^{00}$, like in electrodynamics, is always non-negative for any distribution of the field source in space ($a_0 > 0$):
\begin{equation*}
	T^{00}=a_0\Big\{{1\over c^2}\left({\partial \Psi\over \partial t}\right)^2
	+\left({\partial \Psi\over \partial x}\right)^2+\left({\partial \Psi\over \partial y}\right)^2
	+\left({\partial \Psi\over \partial z}\right)^2\Big\}\geq0\;.
\end{equation*}

The amount of dilaton field energy $dI$ emitted by the source per unit time through a unit solid angle $d\Omega$ is by definition \cite{landau} equal to:
\begin{equation} \label{dI_domega_W}
	\frac{dI}{d\Omega}=\lim\limits_{r\to \infty}r ({\bf W} \cdot {\bf r})\;,
\end{equation}
where ${\bf W}$ is energy flux density vector of dilaton generation. Taking into account that ${\bf W}$ is related to the energy-momentum tensor $T^{ik}$ \eqref{T_ik} as follows: $W^{\alpha}=T^{0\alpha}$, the expression \eqref{dI_domega_W} will be expressed as:
\begin{equation} \label{dI_domega_formula}
	\frac{dI}{d\Omega}=-2a_0\lim\limits_{r\to \infty}r
	( {\bf r}\cdot  \nabla \Psi)\frac{\partial \Psi}{\partial t}.
\end{equation}

Taking into account the expression \eqref{dI_domega_formula}, the angular distribution of the intensity of dilaton generation during the propagation of the magnetic dipole radiation of the pulsar in the Coulomb field of charged particle, averaged over the period of the electromagnetic wave, is equal to:
\begin{align} \label{Q-dil_dI_dOmega}
	\frac{d\bar{I}}{d\Omega}(\theta, \phi) = \frac{8{\cal K}^2a_1^2cQ^2k^4m^2\sin^2\alpha}{a_0r_0^2(1+\cos\gamma)^2}\big[\cos^2\theta_0 \sin^2\theta + \sin^2\theta_0 \cos^2\theta \nonumber \\ - \frac{1}{2}\sin(2\theta_0)\sin(2\theta)\cos(\phi-\phi_0)\big] \big[1-\cos(kr_0(1+\cos\gamma))\big]\;.
\end{align}

It is interesting to note that in the directions of the neutron star's magnetic dipole radiation forward ($\theta =\pi -\theta_0 ,\phi =\phi_0 +\pi$) and backward ($\theta =\theta_0 ,\phi =\phi_0 $), the expression \eqref{Q-dil_dI_dOmega} refers to $0$. In other words, there is no dilaton generation in the forward and backward directions.

Consider the case of $kr_0\ll 1$. Then the expression \eqref{Q-dil_dI_dOmega} will take on a simpler form:
\begin{align} \label{dI_dOmega_apr}
	\frac{d\bar{I}}{d\Omega} = \frac{4{\cal K}^2a_1^2cQ^2k^6m^2\sin^2\alpha}{a_0}\big[\cos^2\theta_0 \sin^2\theta + \sin^2\theta_0 \cos^2\theta \nonumber \\ - \frac{1}{2}\sin(2\theta_0)\sin(2\theta)\cos(\phi-\phi_0)\big]\;.
\end{align}

The expression \eqref{dI_dOmega_apr} has 2 maximums in directions perpendicular to the vector ${\bf r}_0$ \eqref{r_0}, which specifies the location of the charged particle:
\begin{align*}
	\theta = \theta_0 +\frac{\pi}{2},\; \phi = \phi_0\;, \nonumber \\
	\theta = \frac{\pi}{2}-\theta_0, \; \phi = \phi_0+\pi\;.
\end{align*}

By integrating the expression \eqref{Q-dil_dI_dOmega} over the angles $\theta$, $\phi$, one can get the total intensity of dilaton generation averaged over the period of the electromagnetic wave:
\begin{align} \label{tot_I}
	\bar{I} = \frac{8\pi{\cal K}^2a_1^2cQ^2k^4m^2\sin^2\alpha(2-\sin^2\theta_0)}{a_0r_0^2} \nonumber \\\times\big[2(\mathrm{Cin}(2kr_0)-1) + \frac{1}{kr_0}\sin(2kr_0)\big] \;,
\end{align}
where $\mathrm{Cin}(x)$ is the cosine integral defined as:
\begin{equation} \label{Cin}
	\mathrm{Cin}(x)=\int_0^x\frac{d\xi}{\xi}(1-\cos\xi)\;.
\end{equation}
Consider special cases.
The function $\mathrm{Cin}(x)$ \eqref{Cin} has the following asymptotic behavior:
\begin{align*}
	\mathrm{Cin}(x) \approx \frac{x^2}{4} \;,  \;\;x \ll 1 \;, \nonumber\\
	\mathrm{Cin}(x) \approx {\cal C} + \ln x\;, \;\;x \gg 1\;,
\end{align*}
where ${\cal C}=0.577215...$ is Euler`s constant. Hence, when $ kr_0\ll 1 $, the expression \eqref {tot_I} will appear as:
\begin{equation} \label{I_kr0<1}
	\bar{I} =  \frac{16\pi{\cal K}^2a_1^2cQ^2k^6m^2\sin^2\alpha(2-\sin^2\theta_0)}
	{3a_0}\;,
\end{equation}
and when $kr_0\gg 1$:
\begin{equation}\label{I_kr0>1}
	\frac{16\pi{\cal K}^2a_1^2cQ^2k^4m^2\sin^2\alpha(2-\sin^2\theta_0)}
	{a_0r_0^2}\big[{\cal C}-1 +\ln(2kr_0)\big]\;.
\end{equation}
In the expression \eqref{I_kr0<1} one should take into account that $\sin x\sim x -\frac{1}{6}x^3$ at $x\ll 1$.

\section*{Conclusion}

The dilation generation during the propagation of magnetic dipole radiation of a rotating neutron star in a Coulomb field of charged point particle has been considered.
An exact solution \eqref {Q-psi-sphr} of the dilation field equation \eqref {Q_psi_eq_E0Ew} under approximation $|\Psi |\ll 1$ has been obtained.

The intensity of generation of dilatons \eqref{tot_I}, averaged over the period of the electromagnetic wave, has been determined.

It is shown that at $kr_0\gg 1$ the intensity decreases as $r_0^{-2}$ \eqref{I_kr0>1}, and at $kr_0\ll 1$  \eqref{I_kr0<1} does not depend on the distance of the point charged particle from the neutron star $r_0$.

\section{Acknowledgements}
This study was conducted within the scientific program of the National Center for Physics and
Mathematics, section $ \#5$ $<<$Particle Physics and Cosmology$>>.$ Stage 2023-2025.


\begin{thebibliography}{25}
\bibitem{arions-from-pulsar}
Denisov, V. I., B. D. Garmaev, and I. P. Denisova. "Radiation of arions by electromagnetic field of rotating magnetic dipole." Physical Review D 104.5 (2021): 055018.
	
\bibitem{arions-gltc-magn-fld}
Denisov, V. I., et al. "Arions Generation by Magnetodipole Waves of Pulsars and Magnetars in a Constant Magnetic Field." Gravitation and Cosmology 30.2 (2024): 160-164.
	
\bibitem{aroin-3}
Denisova, Irina Pavlovna. "On a mathematical problem in the theory of Goldstone Bosons." Russian Mathematics 64 (2020): 73-77.
	
\bibitem{axion-1}
Cheng, S. L., C. Q. Geng, and W-T. Ni. "Axion-photon couplings in invisible axion models." Physical Review D 52.5 (1995): 3132.
	
	
\bibitem{axion-2} 
Klimchitskaya, G. L., P. Kuusk, and V. M. Mostepanenko. "Constraints on non-Newtonian gravity and axionlike particles from measuring the Casimir force in nanometer separation range." Physical Review D 101.5 (2020): 056013.

\bibitem{axion-3}
Patkos, Andras. "Electromagnetic energy transfer processes in effective electro-magneto dynamics of axions." Modern Physics Letters A 38.30n31 (2023): 2350137.
	
\bibitem{dilaton-1}
Fahim, Bardia H., and Masoud Ghezelbash. "New class of exact solutions to Einstein?Maxwell-dilaton theory on four-dimensional Bianchi type IX geometry." The European Physical Journal C 81 (2021): 1-17.

\bibitem{dilaton-2}
Ripley, Justin L., and Frans Pretorius. "Scalarized black hole dynamics in Einstein-dilaton-Gauss-Bonnet gravity." Physical Review D 101.4 (2020): 044015.
	
\bibitem{dil-from-pulsar}
Denisov, V. I., I. P. Denisova, and E. T. Einiev. "The investigation of low-frequency dilaton generation." The European Physical Journal C 82.4 (2022): 311.
	
\bibitem{dil-Q-plane_wave}
Denisova, I. P. "Dilaton Generation While a Plane Electromagnetic Wave Propagates in Coulomb Field." Gravitation and Cosmology 27 (2021): 392-395.
	
\bibitem{string-th}
Youm, Donam. "Black holes and solitons in string theory." Physics Reports 316.1-3 (1999): 1-232.

\bibitem{Kaluza-klein}
Overduin, James Martin, and Paul S. Wesson. "Kaluza-klein gravity." Physics reports 283.5-6 (1997): 303-378.
	
\bibitem{ALP-exp1}
Banerjee, D., et al. "Search for axionlike and scalar particles with the NA64 experiment." Physical review letters 125.8 (2020): 081801.
	
\bibitem{ALP-exp2} 
Inada, T., et al. "Search for two-photon interaction with axionlike particles using high-repetition pulsed magnets and synchrotron X rays." Physical Review Letters 118.7 (2017): 071803.


\bibitem{landau}	
L.D.Landau, E.M.Lifshitz, {\it The classical theory of fields},
(Butterworth-Heinemann, 1975).

\bibitem{NS_rad}
Capano, Collin D., et al. "Stringent constraints on neutron-star radii from multimessenger observations and nuclear theory." Nature Astronomy 4.6 (2020): 625-632.


\end{thebibliography}
\end{document}